# Electronic Structure of Multiple Dots

M. Stopa[a], A. Vidan[a], T. Hatano[b], S. Tarucha[b,c] and R. M. Westervelt[a]

[a]Center for Nanoscale Systems, Harvard University, 17 Oxford St. Cambridge, MA 02138
[b] Quantum Spin Information Project, ICORP, Japan Science & Technology Agency, Morinosato Wakamiya 3-1, Atsugi-shi, Kanagawa, 243-0198, Japan ERATO
[c]Department of Applied Physics, University of Tokyo, Hongo, Bunkyo-ku, Tokyo, 113-0033, Japan

We calculate, via spin density functional theory (SDFT) and exact diagonalization, the eigenstates for electrons in a variety of external potentials, including double and triple dots. The SDFT calculations employ realistic wafer profiles and gate geometries and also serve as the basis for the exact diagonalization calculations. The exchange interaction J between electrons is the difference between singlet and triplet ground state energies and reflects competition between tunneling and the exchange matrix element, both of which result from overlap in the barrier. For double dots, a characteristic transition from singlet ground state to triplet ground state (positive to negative J) is calculated. For the triple dot geometry with 2 electrons we also find the electronic structure with exact diagonalization. For larger electron number (18 and 20) we use only SDFT. In contrast to the double dot case, the triple dot case shows a quasi-periodic fluctuation of J with magnetic field which we attribute to periodic variations of the basis states in response to changing flux quanta threading the triple dot structure.

The spin state of electrons in multiple quantum dot assemblies formed in two dimensional electron gas (2DEG) semiconductor heterostructures is determined by exchange interactions between the electrons and not, generally, by the much smaller magnetic dipole interactions between the spins. In the simplest case of two electrons in two dots (artificial molecular hydrogen) competition between exchange, which favors spin alignment (spin triplet), and tunneling, which delocalizes the electrons and tends [1] to favor spin anti-alignment (spin singlet) can be modulated by a magnetic field and is sensitive to the precise geometric nature of the tunnel barrier separating the two dots [2]. The exchange splitting J is defined as the energy difference between the ground state triplet and the ground state singlet (and is distinct from the exchange integral which is one of the Coulomb matrix elements between two-electron states) and is crucial to the implementation of various schemes of quantum computation [3].

In this paper, we study the spin state of double and triple quantum dots (in a ring configuration) as a function of electron number N, magnetic field B, and the various gate voltages and gate geometries controlling the height and shape of the barriers between the dots. We briefly describe results of N = 2 exact diagonalization calculations in a realistic double dot geometry (more details for this case can be found in Ref. [4]) and then focus on the triple quantum dot. This latter case we explore in two regimes with two methods. First, we extend the N = 2 exact diagonalization method to the three-dot case. Next, we consider the case of many electrons calculated within spin density functional theory (SDFT). The triple dot stability diagram [5,6] identifies, in particular, the N = 20 case as similar to the N = 2 case. Specifically, since 18 electrons constitute a filled shell (for the triple dot), N = 20 has two valence electrons. We first examine the structure of the magneto-spectrum for the "filled-shell" N = 18 case. It is then of interest to study how the 18 core electrons influence the electronic structure of two valence electrons when we increase N to 20.

Here, we consider the case where the three dots are as nearly "balanced" as possible. This is straightforward for N = 18, where 6 charges occupy each dot and a spectral gap exists to the next (empty) state. For N = 20, however, it is necessary to set the gate voltages carefully so as to maintain a charge of 20/3 in each dot. Likewise, for N = 2 we maintain the bias such that 2/3 of an electron is resident in each dot. We show that the effective single particle spectrum (the Kohn-Sham levels) evolves, as a function of applied magnetic field B, in the form of a triply degenerate (six-fold degenerate if spin is included) Fock-Darwin spectrum [7]. The splitting between the three states can be calculated with perturbation theory at B = 0. In addition we examine the exchange energies for the N = 2 triple dot as a function of B. It is well-established that for a double dot J(B) is positive for B = 0 but can cross to negative (triplet ground state) as B increases before finally saturating to zero as the magnetic field effectively isolates the two dots. For the triple dot we show that J(B) oscillates between positive and negative several times before decaying to zero. This suggests a competition between the exchange integral and the tunnel coupling that is mediated by the evolving single particle states which are in turn influenced by the magnetic flux that threads the triple dot device.

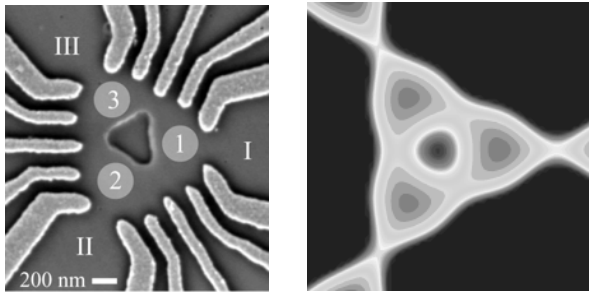

**FIGURE 1.** (left) SEM of triple dot device illustrating multiple surface gates and etched central region, (right) self-consistent potential profile of structure, potential cutoff at 1 Ry* ≈ 5.8 meV above the Fermi surface.

We calculate the electronic structure of GaAs-AlGaAs modulation doped heterostructure-based quantum dots formed via standard split gate techniques [8], see Fig. 1. Our calculations of the electronic structure of such dots and multiple dots performed with spin density functional theory (SDFT) have been documented extensively in the literature [9]. Note that here we are describing the results of two different types of calculation. The first uses the standard SDFT method. The second method, which we apply to 2 electron dots (or multiple dots) only, is an exact diagonalization calculation which employ the SDFT results (the Kohn-Sham states) as a basis only. This method, which we will describe in detail in a future publication [4], incorporates the full geometric fidelity of the structure while also including the full effects of many-body correlation. It is, of course, limited to small particle numbers.

Numerous recent calculations (see Fig. 4a), motivated by the quantum computation implications of manipulating the exchange interaction, have demonstrated that for N = 2 double dots (artificial molecular hydrogen) the singlet nature of the B = 0 ground state gives way to a transition to a triplet ground state at finite B [1,10]. Physically, spin-alignment (triplet) is favored by the exchange integral, as in the case of Hund's rules for atoms, and spin anti-alignment (singlet) is favored by tunneling which permits double-occupancy through delocalization. We have recently shown that the action of the magnetic field is to compress the wavefunction overlap in the saddle point between the two dots [4]. This enhanced overlap increases exchange (integral) while leaving the tunneling coefficient essentially unaffected. This therefore constitutes an explanation of the B-dependent crossover of J. For triple dots, the behavior of J(B) is more complicated. Prior to discussing these N = 2 results, however, we describe the SDFT results for triple dots with many electrons.

In Fig. 2 we show the calculated energy levels (SDFT) for a single spin species ("spin up") for a triple dot with N = 18 and N = 20 as a function of magnetic field. Note that the dot is strongly isolated from the leads and that the Fermi energy of the leads (which is the energy zero of the problem) does not coincide with the Fermi surface, $E_F$, of the dot, which for N = 18, lies between the 9th and the 10th levels (10th and 11th for N = 20). Points are calculated every 0.1 T. The resemblance to the Fock-Darwin spectrum is evident. The tunnel coupling between the dots is small (it is of the order of the splitting between the sets of three states at B = 0). Thus the spectrum is predominantly that of three isolated dots, each approximately parabolic, nearly degenerate except for a small tunnel splitting. It is interesting that there is no evident effect of the flux threading the dot ring in these results. For N = 20 the spectrum is quite different due to the location of the dot Fermi level within a group of three (six, including spin) levels. The nearly degenerate levels are influenced by small changes in B which further affects the occupancies. These shifts in occupancy thereupon change the self-consistent potential. So, while the gates are set at B = 0 to "balance" the three dots, the magnetic field breaks this symmetry and produces charge redistribution. An additional effect, which is an anomaly of mean field theory, is that levels tend to stick together (unless there is a gap at the Fermi surface) so that fractional occupancy can produce a charge distribution that best minimizes the electrostatic energy.

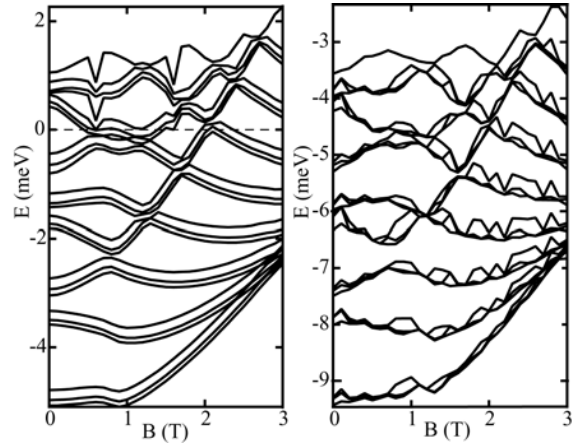

**FIGURE 2.** Kohn-Sham energy levels for triple dot (spin up) with (left) N = 18 electrons and (right) N = 20 electrons. For N = 18, each electron has a filled Fock-Darwin "shell" of 6 electrons and so the Fermi level is in a gap. For N = 20, fluctuations of the self-consistent structure occur as B slightly modifies the Fermi surface states. Points calculated every 0.1 T.

In order to reduce the electron number to N = 2 it is necessary to make the surface gate voltages considerably more negative and shrink the area of the dots. (In practice, this could make them inaccessible from the leads). The Kohn-Sham level structure reflects this smaller size, as seen in Fig. 3 where the spacing between the lowest three states and the next six (s and p, respectively, in the Fock-Darwin spectrum) is greater than in the larger N cases.

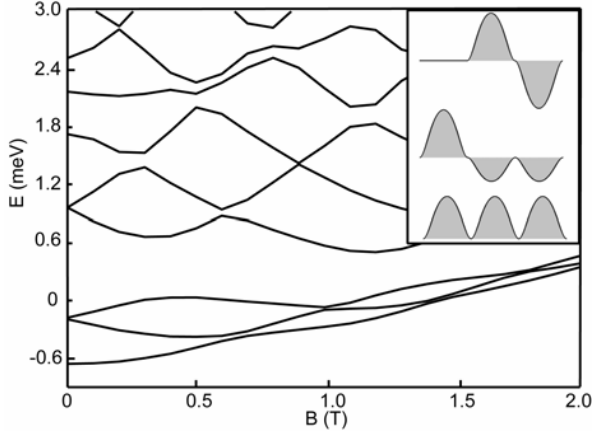

**FIGURE 3.** Kohn-Sham energy levels for the N = 2 triple dot. Splitting, at B = 0, between first level and second pair is 3t. Inset shows single particle levels of three dots in a ring in tight binding model. Note that level 2 fluctuates between levels 1 and 3 with the same behavior as the fluctuation of J seen in figure 4.

In order to understand the level structure, it is instructive to diagonalize the single particle Hamiltonian for the triple dot in the tight-binding basis, where

$$H = \begin{pmatrix} \varepsilon & -t & -t \\ -t & \varepsilon & -t \\ -t & -t & \varepsilon \end{pmatrix},$$

assuming that the dots are identical with energy $\varepsilon$ and connected with tunneling coefficient t. The eigenvalues of H are $E = \varepsilon - 2t$ and $E = \varepsilon + t$, the latter of which is doubly degenerate. The (un-normalized) eigenfunctions, shown in the inset to Fig. 3, are: (1,1,1), (1,-1/2,-1/2) and (0,1,-1). Therefore, the splitting between the lowest level in Fig. 3 and the ensuing degenerate pair, ~0.4 meV, is 3t.

As the magnetic field is turned on, level fluctuations that are not apparent in Fig. 2 begin to occur. We attribute these fluctuations, most notably the oscillation of level 2 back and forth between levels 1 and 3, to the addition of flux quanta through the ring formed by the three dots. A major consequence of this variation is the behavior of the singlet-triplet splitting J(B) shown in Fig. 4b. Note that in contrast to the double dot case, whose representative calculations of J(B) are illustrated in Fig. 4a, the exchange begins positive and then changes back and forth between positive and negative several times.

A basic explanation of the fluctuating J(B) is as follows. The lowest state in Fig. 3 can contain two electrons, due to spin degeneracy. Where the second orbital descends to near degeneracy with the first (at B ~ 0.6 T, for example) the additional orbital degeneracy allows spin alignment, in the form of Hund's rule. Therefore in these regimes the triplet is favored. While a more thorough explanation in terms of the competition between delocalization and exchange is desired, it is clear that the driving mechanism for J(B) oscillation is the influence that the threaded magnetic field exerts on the evolving basis states.

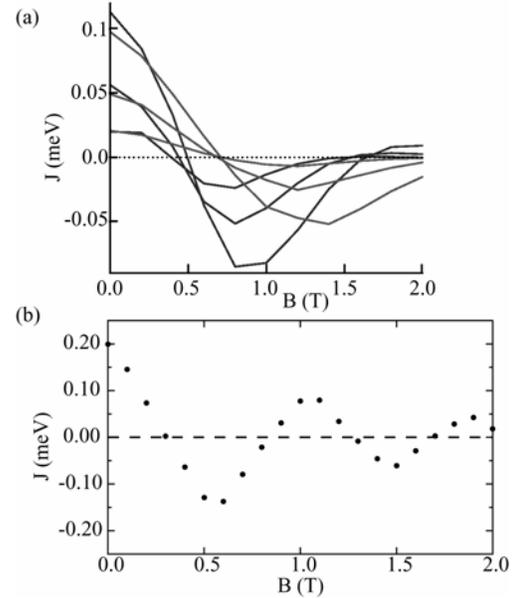

**FIGURE 4.** Exchange (singlet-triplet splitting) as a function of magnetic field for (a) double and (b) triple dot calculated from full exact diagonalization method for N = 2. Note that, in contrast to the double dot case (plotted here for a typical variety of gate patterns and tunnel coupling strengths), J(B) for the triple dot exhibits several oscillations while it is damped by increasing localization of the eigenstates within the individual dots by the magnetic field.

In conclusion, we have used density functional theory and exact diagonalization, formulated on a density functional basis, to investigate the exchange energy in double and triple quantum dots. We find that the single particle spectrum for many (~20) electrons has the form of a triply degenerate Fock-Darwin spectrum, with little influence of the threading magnetic field. We have also shown an oscillatory structure to J(B) for the triple dot in a ring which contrasts strikingly with that of the double dot case and which has not heretofore been discussed.